# Monte Carlo modeling photon-tissue interaction using on-demand cloud infrastructure

Ethan P. M.LaRochelle[1], Pedro Arce[2], Brian W. Pogue[1]

[1] Thayer School of Engineering at Dartmouth, Hanover NH 03755
[2] CIEMAT (Centro de Investigaciones Energéticas, Medioambientales y Tecnológicas), Madrid, Spain

E-mail: ethan.phillip.m.larochelle.th@dartmouth.edu

## Abstract

**Purpose:** This work advances a Monte Carlo (MC) method to combine ionizing radiation physics with optical physics, in a manner which was implicitly designed for deployment with the most widely accessible parallelization and portability possible.
**Methods:** The current work updates a previously developed optical propagation plugin for GEANT4 architecture for medically oriented simulations (GAMOS). Both virtual-machine (VM) and container based instances were validated using previously published scripts, and improvements in execution time using parallel simulations are demonstrated. A method to programmatically deploy multiple containers to achieve parallel execution using an on-demand cloud-based infrastructure is presented.
**Results:** A container-based GAMOS deployment is demonstrated using a multi-layer tissue model and both optical and X-ray source inputs. As an example, the model was split into 154 simulations which were run simultaneously on 64 separate containers across 4 servers. **Conclusions:** The container-based model provides the ability to execute parallel simulations of applications which are not inherently thread-safe or GPU-optimized. In the current demonstration, this reduced the time by at most 97% compared to sequential execution. The code and examples are available through an interactive online interface through links at:
https://sites.dartmouth.edu/optmed/research-projects/monte-carlo-software/

## Introduction

Stochastic modelling by Monte Carlo (MC) methods are used as the fundamental standard to simulate effects in human tissues. The current work details the extension of previous efforts by our group, by updating MC radiation-optical software and demonstrating the ability to programmatically deploy multiple copies of a MC software package onto a cloud-based infrastructure for efficiently running complex simulations. While these updates have been packaged as a virtual machine (VM), this paper describes how to apply the concept of containerization to further lower the overhead needed to run each simulation in parallel on a single system, or cluster of systems. GAMOS and the optical plugin can be used with positron or beta emitting radiotracers,[1,2] or external beam therapy,[3–5] as a tool to generate optical excitation and emission signals within tissue. While cloud-based MC packages,[6–8] and the concept of lowering system overhead in high performance computing (HPC) using containers have both been proposed previously,[9,10] the current report provides a distributable platform for simulating both high-energy physics and optical transmissions on a cloud platform utilizing a container architecture, allowing for scalable parallelization. Extensive supplemental information is available providing access to a VM (S1), the optical plugin code modifications for GAMOS 6.1 (S2), validation examples (S3), interactive code used to generate the container and cloud infrastructure (S4) and the final container image (S5). The goal of this work is to document how an MC platform can be packaged such that it can be distributed and executed in a parallel manner without introducing underlying code changes to ensure thread-safety or GPU optimization.

This work relies on the concept of a container, which acts much like a VM with lower resource requirements. Containers provide semi-isolated execution environments composed of a user-defined set of packages and read-only access to the underlying libraries and services on the host operating system, which drastically reduces their footprint compare to VMs. A single script file is used to define the container environment which can be stored as an image on a remote repository. Multiple containers can then be programmatically installed on a single or multiple host systems.

Estimating optical light propagation through heterogeneous turbid media is often addressed through MC methods, where many of the current software solutions stem from the original work by Wang and Jacques who developed the publicly available Monte Carlo for Multi-Layer media (MCML),[11] which is used to approximate photon fluence in layered tissues. Since the development of MCML, many other optical MC packages have been developed, each solving a specific problem: Monte Carlo XYZ (MCXYZ) uses voxelized optical properties,[12] Monte Carlo eXtreme (MCX) also uses voxelized geometry but is optimized for GPU acceleration,[13] Mesh-based Monte Carlo (MMC) uses a mesh-based geometry to reduce computational complexity,[14] while MCflour is another example which focuses on modelling fluorescence but is limited to homogeneous media.[15] A number of other optical MC packages exist, with many recent publications focusing on optimization of computation time through GPU,[16–18] or parallel execution.[19–21]

The use of MC methods is widely adopted in the field of high energy physics, where Geometry and Tracking (Geant4) software, maintained by an international consortium, is a core resource used by most in the field.[22–24] While Geant4 is widely used, it also has a steep learning curve, requiring the user to write C++ code to define world geometries, input events, and detector processes. In the field of medical physics, the need to write and compile C++ code has been addressed through Geant4 architecture for medically oriented simulations (GAMOS), which is a software package designed to interface with many of the Geant4 features using simple script files which do not need to be compiled by the user.[25] A plug-in for GAMOS to account for optical photon transport, including tissue optical properties and Cherenkov emissions was previously developed and validated by Glaser et al and our current work used the same validation methods before demonstrating the cloud-based deployment functionality.[26]

A diagram showing the relation of Geant4 to GAMOS is provided in Figure 1A, where the Geant4 kernel has three main components: 1) the underlying physics model, 2) the processes to define the inputs and track the events over multiple steps, and 3) a geometry definition which considers the placement and composition of every aspect of the simulation. GAMOS wraps this implementation so the functionality of Geant4 can be called using a number of plug-ins. A command interpreter is used to convert two text-based input files: the command script, and geometry file, which are used to describe all parameters necessary to run a simulation. GAMOS is also configured to utilize ROOT, a software package developed at CERN for visualizing large datasets.[27] While ROOT is not a core component of GAMOS, it is part of the installation package and, as shown in Figure 2.1-1B, accounts for 3.8 million lines of code which is 4X more code than the core dependency Geant4. An analysis of the code structure of Geant4 shows the relative distribution of the code (Figure 1C), where processes such as managing transport for different physics lists, accounts for the majority of the logic.

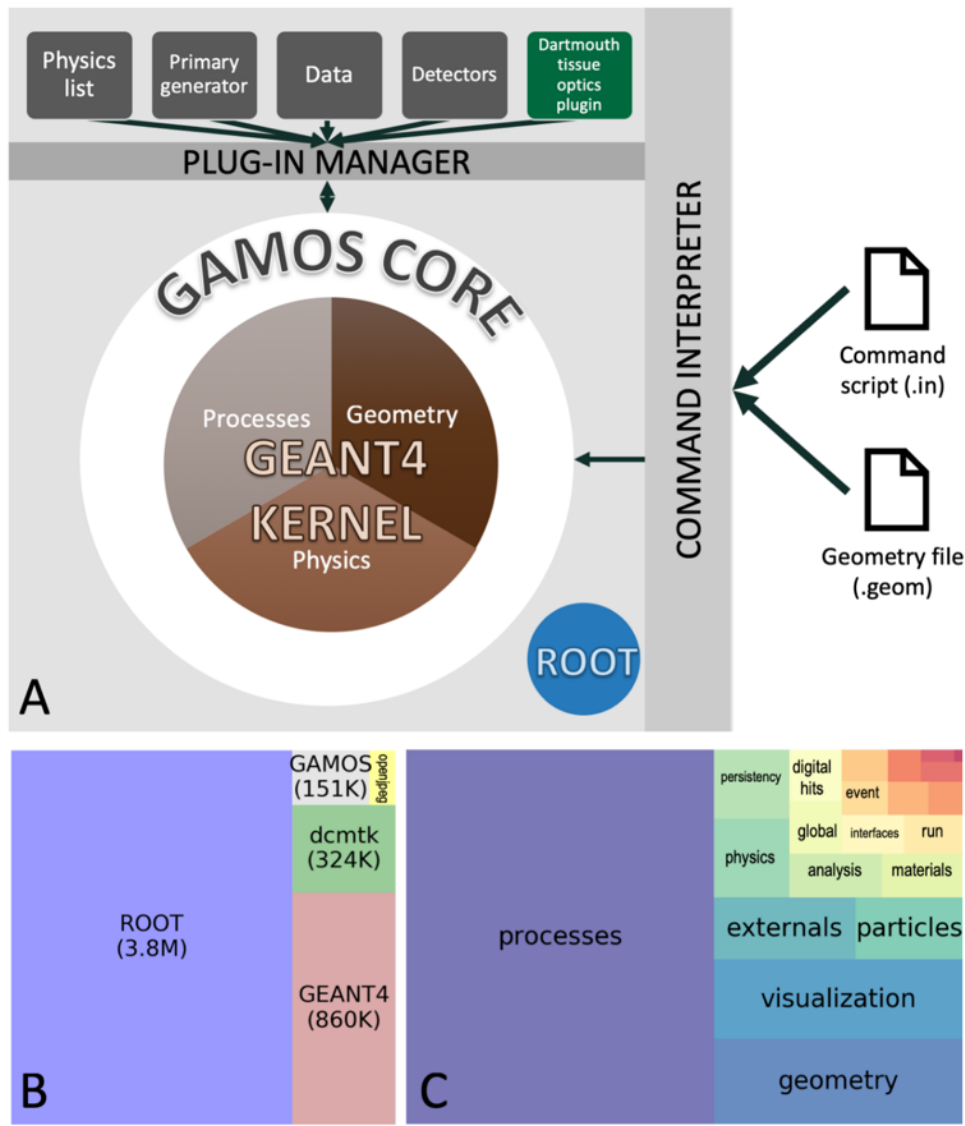

*Figure 1: GAMOS architecture relies on Geant4 and allows for user-defined plug-ins (A). GAMOS is installed with external dependencies, some of which are larger than the base program (B). The complexity of the Geant4 code-base is shown for each sub-process, where the area of each box represents the proportion of the number of lines of code. (C)*

The optical-photon transport functionality was designed as a plug-in to GAMOS which allows users to attach material properties to define wavelength-specific absorption, scattering, anisotropy, refractive index and fluorescence properties.[26] Additionally, routines for generating Cherenkov and scintillation emissions were implemented. These changes allowed the ability to run MC simulations involving high-energy physics and tissue optical properties. However, since the tissue optics plugin for GAMOS was first published, GAMOS has transitioned from version 4.0.0 to 6.1.0. As a result, much of the functionality of the original plugin suffered from broken dependencies. While other X-ray-optical interaction MC packages exist,[28–30] many lack the full functionality provided by the Geant4-GAMOS kernel. As part of the current work, the original functionality of the plugin has been restored by addressing the broken dependencies, where all changes are publicly available for review using code versioning (S2), and validation scripts provided with the original publication were used to verify functionality (S3). These modifications have been packaged into both a virtual machine (S1) and a low-overhead container architecture (S4, S5). There is a growing trend in MC packages to utilize a combination of multi-threading, GPU optimization and parallel execution to optimize execution time.[19–21,6,8] While GAMOS is not designed for multi-threaded or GPU optimization, we present a method to achieve parallel execution. A demonstration of programmatically deploying multiple cloud-based instances of GAMOS 6.1.0 with the tissue optics plugin is provided in the supplementary material (S4), where multiple simulations were conducted in parallel to track optical emissions of a tumor model for both optical and X-ray input sources.

## Materials and Methods

The latest version of GAMOS (6.1.0) was downloaded to a virtual machine running Ubuntu 18.04.03 LTS. A version control system (git) was used to track all changes made to the GamosCore source directory (S2). The files added for the tissue optics plugin in GAMOS version 4.0.0 were re-added to their respective sub-directories. Modifications to the appropriate headers and plugin settings were made in the source files. Generally, the overall operation of GAMOS at a function-level remained unchanged. The modified GAMOS was then re-compiled using the provided installation scripts. All validation scripts used for the original tissue optics plugin and available in the supplementary material (S3) were then run to verify appropriate functionality.

A full GAMOS installation requires approximately 7.4GB of storage, as shown in Figure 2A. Since the development was performed on a virtual machine, a snapshot of the system can be exported to a single portable file using the Open Virtualization Format. This was performed using VirtualBox, resulting in an 8.55GB .ova file, which includes the application and the guest OS (Available at S1).

As an additional delivery mechanism, the code was also packaged in a container using an application called Docker.[31,32] Docker uses a single-file script (Dockerfile) to programmatically define a system image which it places in a container, which acts much like a virtual machine with lower resource requirements. The download scripts commonly used to install GAMOS were retrieved using the wget command. The getGamosFiles.sh script was modified to add commands to replace the GamosCore source directory with the one containing the tissue optics plugin and the corresponding validations.  A file used to install additional package dependencies

(installMissingPackages.Docker.Ubuntu.18.04.sh), mainly required by ROOT, was also placed in this directory. An additional script provided to interface with Amazon Web Services (AWS) Batch[33] (fetch_and_run.sh) was also placed in this directory.

A text-based Dockerfile (provided in S4) is used to define the container specifications and is built on top of an Ubuntu 18.04 image, much like our initial virtual machine. The file is used to install GAMOS and update the installation with the tissue optics plugin. It also configures the container to read and write using a cloud-based storage service.

Once the Dockerfile sufficiently describes the desired system parameters the docker build command is used to create the container. This runs the Dockerfile script to create the application image which can be referenced by multiple containers on the same system without interfering with each other or installing multiple guest operating systems like in a (Figure 2B-C). The container can then be stored in a repository so any system running Docker can retrieve the application image and run one or many containers. In the present example the container repository is AWS Elastic Container Repository (ECR) but the same file is publicly available on DockerHub, linked in Supplementary Material (S5). The newly built container is tagged with a name and pushed to this repository for later retrieval.

Using a Python API interface for AWS (Boto3),[34] the infrastructure needed to access and configure these containers as well as define parameters for parallel execution can be defined. An interactive notebook with all code needed to configure the environment is available in the Supplementary Material (S4).

Using the AWS web interface, a new user with *AdimPowerUser* group security settings can be created and assigned an API token. The Boto3 API will use these configuration settings which can be stored on the local computer with read-only user permissions. Storing the access keys in a file improves security by reducing the chance the secret token is displayed and logged when executing API commands, and also reduces the need to manually specify the security credentials with every API call. Once the local Boto3 package is configured, it can be used to create and modify the infrastructure described in Figure 3.

To implement large parallel executions of GAMOS, AWS Batch was leveraged to define the computational infrastructure and job queue.[33] Batch provides an interface between the Elastic Container Service (ECS), compute resources (EC2), and storage services (S3) to efficiently replicate multiple container images and assign tasks in an on-demand basis.

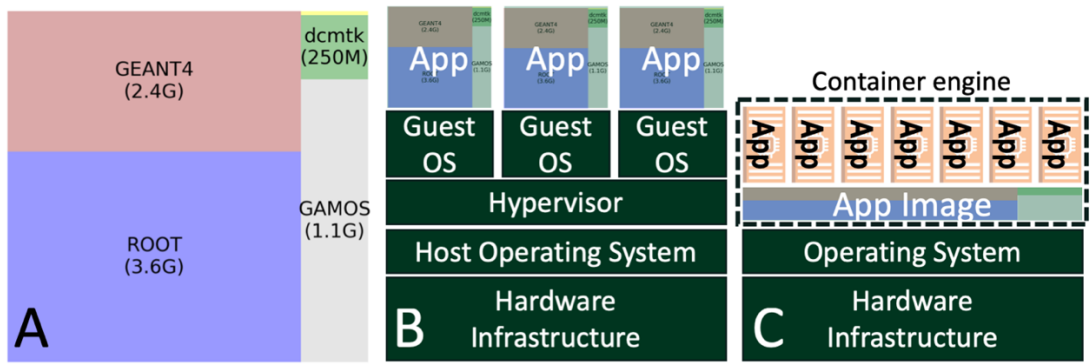

*Figure 2: The relative size of GAMOS and the external packages are shown (A). The concept of how the GAMOS installation is deployed using a Virtual Machine (B) or Container (C) architecture.*

Since GAMOS is a relatively large application for a container, a launch template was used to expand the basic container storage specification. The launch template also specified the default server image and networking parameters used when initializing a new compute resource. With this launch template in place, Batch is used to configure a computing environment, which specifies the total number of virtual CPUs (vCPU) and instance types.  The instance types are used to specify the size and specialty of the requested server, where they can be optimized for compute, memory, GPU, etc. In the present example the instance types were limited to EC2 instance types c5.2xlarge, c5.4xlarge, c5.9xlarge, which are the names given by AWS, where the first letter indicates compute-optimized instances, the 5 indicates the hardware is an AWS 5th-generation system, and the size is correlated with the number of vCPUs and system memory available to the system. In this case a c5.2xlarge has 8 vCPUs and 16GB of memory, whereas the c5.9xlarge has 36 vCPUs and 72GB of memory. A maximum of 128 vCPUs was set for this environment which are automatically allocated by AWS Batch when submitting jobs. Additionally, this service is used to automatically stop instances when the queue is empty, after jobs are finished.

A Batch job queue was then defined and given a priority, where the priority is used when multiple queues are active. The previously defined compute environment was added as an available resource associated with this job queue. Once the compute environment and job queue are defined, individual jobs can be created and submitted to the queue. All jobs are given a unique name and provided with specific container and environment properties and then sent to a queue for processing.  The container properties define the location of the container repository and desired container image, as well as what resources to assign to the container. In this case 2 vCPUs and 2 GB of memory were allocated to each container, where 2 vCPUs were chosen because this is the smallest number used by an EC2 instance. The container environment settings then specify the location of a compressed zip file on the network-accessible Simple Storage Service (AWS S3), and also provide input arguments. The input arguments specify the shell script, contained in the zip file, to execute and additional input arguments specifying a random seed, number of events, and other simulation-specific parameters. Since the job definition is created using a Python script, it can be wrapped in a loop to specify a range of input parameters and submit the job to the queue for processing. The zip file also provides templates of GAMOS input files and any necessary extra inputs needed to run the simulation.

After the job queue becomes populated with jobs, Batch verifies the validity of the environmental parameters. The compute autoscaling service then automatically requests resources, which are loaded with the specified container image (Figure 2C, Figure 3), and the container is replicated based on the vCPU and memory allocations defined by the job. For example, since c5.9xlarge instance have 36 vCPUs and 72 GB of memory, and our jobs were allocated to containers with 2 vCPUs and 2 GB of memory, the number of containers on an instance is limited by the number of vCPUs, or 18 containers on this instance.

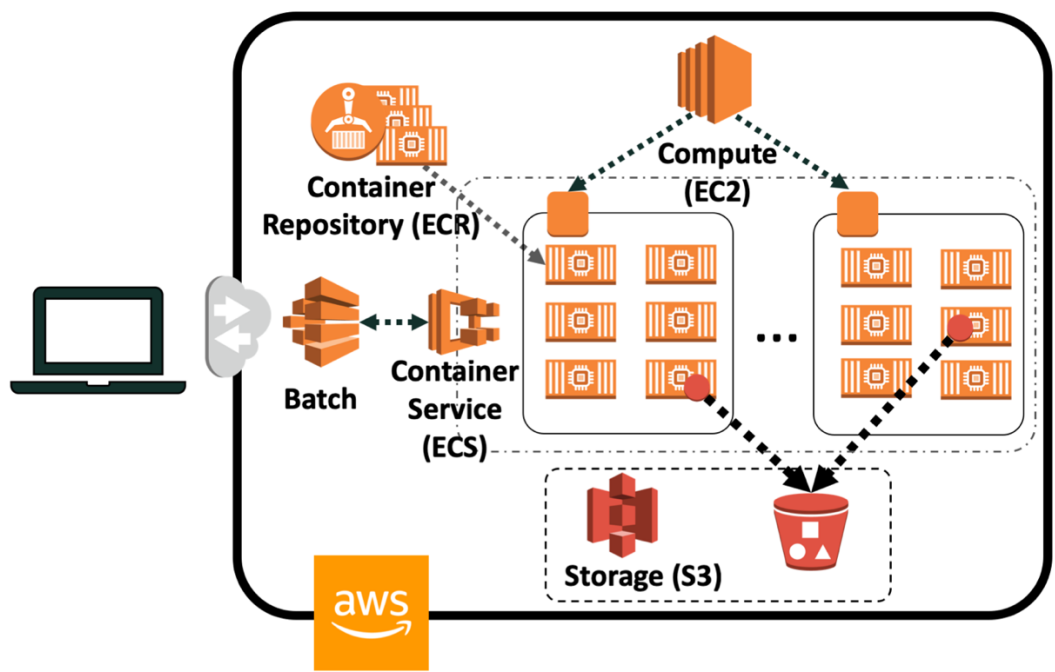

*Figure 3: A programmatically configured cloud-infrastructure can be used to send jobs to individual containers located on multiple servers, which send results to specific storage buckets.*

Each container executes the code defined by the provided shell script, using unique input arguments provided in the job definition. The shell script is used to modify the GAMOS input file templates provided to the container and run the GAMOS simulation. While this script runs, the status can be monitored from the AWS CloudWatch web interface, where all standard terminal outputs for each container are logged in real-time. The bash script also contains commands to copy the simulation directory to a container-specific sub-folder of the AWS S3 storage bucket containing the original simulation definition zip file. So, after a set of jobs is complete, all results are stored in sub-folders of a single AWS S3 bucket which can be programmatically retrieved for analysis.

## Results

Briefly, the container-based cloud deployment of GAMOS was used to compare the fluorescence emissions of a tumor inclusion and light fluence in a multi-layer tissue model, using a total of 154 simulations provided in supplementary material (S4). The number of interactions that are tracked for each event governs the speed at which these simulations can be run, where the run-time of each of the 154 simulations was recorded and averaged for each of the four source types, as depicted by the average system time in Table 1. In this model the tumor is given tissue optical properties with the addition of fluorescence absorption similar to that of PtG4, a platinum porphyrin with absorption at approximately 430nm and 630nm and emission peak near 770nm.[35] Since 430nm light is highly attenuated by skin, the average execution time of these simulations is on the order of minutes, whereas deeper penetrating 630nm light will propagate further through tissue and more event interactions are tracked, resulting in a longer execution time. When high-energy X-rays are used, multiple optical photons are generated through Cherenkov emissions along the path of scattered high-energy electrons, where each of these secondary emissions are then tracked, further increasing the simulation complexity and execution time.

The compute cluster defined for these simulations specified a maximum number of virtual CPUs as 128, which was the limiting factor for container distribution. With 128 vCPUs, 64 containers could run simultaneously. Each simulation has an approximately 5-minute overhead needed to initiate the container infrastructure, but scaling allows for parallel execution which provides the ability to execute all simulations in the same time required to execute a single simulation. With the Cherenkov simulations, 7 tumor inclusion depths were specified, and each depth had 10 simulations of $10^6$ events, each with a different random seed provided as an input argument. Since our compute-environment only specified 64 containers, and the 6MV and 18MV simulations were split into 70 simulations, approximately 90% of these simulations were completed within the average system time required to complete one simulation, but the final 10% required additional time to be executed. However, due to the auto-scaling natures of this system, the final 10% could be finished on a smaller compute resource.

*Table 1: Execution times for GAMOS simulations with given source events*

|  | *430nm* | *630nm* | *6MV* | *18MV* |
|---|---|---|---|---|
| *Depths tested* | 7 | 7 | 7 | 7 |
| *Total simulations* | 7 | 7 | 70 | 70 |
| *Average system time* | 10 min | 55 min | 90 min | 200 min |
| *Sequential time* | 70 min | 6.5 hr. | 105 hr. | 233 hr. |
| *Parallel execution time* | ~15 min | ~1 hr. | ~3 hr. | ~7 hr. |

## Discussion

GAMOS is a powerful Monte Carlo software package well suited for medical physics simulations. It has a large footprint though, which relies on many dependencies, and can make compilation during installation a time-consuming endeavor. While the plugin architecture provided the opportunity for Glaser et al. to expand the capability of GAMOS into the field of biomedical optics, this plugin was not kept current with subsequent GAMOS releases. As such, since only the source code of the plugin was released, when GAMOS was updated, the functionality to use these features was lost. The current work has provided updates to this optical light transport plugin, so it is again compatible with the latest version of GAMOS. Additionally, both a VM and container images of these changes have been made available as a way of removing the need to compile code, the main bottleneck in installation. While efforts will be made to maintain this functionality in future GAMOS releases, in the event this does not occur, the VM and container-based packages will continue to provide needed functionality to users requiring these features.

While other Monte Carlo software packages exist for modeling light transport[6-16] or high-energy physics,[22,36–38] fewer have intersecting capabilities.[23-25] While users may be able to accomplish this through Geant4, it also requires the ability to write sometimes complex C++ code. Other examples of similar software package are the Geant4 wrapper PTSim and TOPAS.[38,39] Much like GAMOS, PTSim and TOPAS provide a similar text-based input structure, targeting particle therapy applications, and currently lack the optical transport plugin provided by Glaser et al.[26] A separate collaboration, Geant4 Application for Emissions Tomography (GATE), has developed similar macro script-based interface for modeling medical imaging applications, specifically targeting movement and temporal-based models,[40] and this base deployment was recently demonstrated in a similar container architechture.[9] There are also optical transport plugins for GATE, which are capable of modeling bioluminescence and florescence.[41] Recent advances to GATE have also demonstrated the ability to simulated and track Cherenkov and scintillation emissions,[42] so much of the same functionality could be achieved with GATE. While each of these packages target specific user communities, their architecture is such that they could be deployed in a similar container-based infrastructure as described here.

There is a growing trend to use rapidly advancing computational infrastructure to further optimize the execution time needed to run MC simulations. Much of this effort has focused on GPU-optimization,[16–18,30,43] however this generally requires the underlying MC packages to be written such that it can leverage these resources. Another approach has been to develop methods for parallel execution on multiple CPUs or cores.[19–21] These methods often focus on developing a task distribution system for available compute nodes, which can be a time-consuming design endeavor. Others have proposed cloud-based execution previously,[6,8,9] which the current work employs by describing a method to achieve parallel execution using Python scripts and an on demand infrastructure. This greatly reduces the capital investment needed to purchase and maintain a dedicated server infrastructure and reduces the need to re-write thread-safe or GPU-optimized code.

Through the development of a container, the overall resources needed to run multiple GAMOS instances is greatly reduced. GAMOS is not inherently thread-safe, so containers provide the ability to compartmentalize the execution of simulations with independent compute resource, all on the same server. While the present example is provided for deployment on a cloud-based infrastructure, with minor modifications it could also be run on a locally managed server or a single institution's computing cluster.

The companion code provided with this publication provides an example of how a cloud-based infrastructure can be deployed programmatically (S4). This could be useful for researchers with limited resources, where a local compute cluster is not available, or a high-end server is impractical. In the provided examples, billing was based mainly on the compute-time required to accomplish, so for example, a simulation utilizing 1 vCPU for 100 hours is billed at a similar rate as 100 vCPUs utilizing 1 hour. While there are limits on the maximum resource which can be utilized at any one time, it may have benefits for users who wish to quickly obtain results, without purchasing a powerful server.

The framework presented by the current work focus on deploying a GAMOS container in a cloud-based infrastructure, but with slight modifications the same framework could be used to deploy any opensource software which can be run from the command line. This provides a simple pathway for other Monte Carlo packages mentioned previously to be run in a similar manner. While the size of the current container was relatively large, many other packages are much smaller can could be easily implemented in a container. It may also be possible to install GAMOS without ROOT, which would reduce the container size by approximately 3.6GB, although this was outside the scope of the current work.

A demonstration of the capability of this platform was conducted by executing over 150 simulations with varying input conditions. If these simulations were run sequentially it would have taken over 14 days of compute time, however using 64 simultaneous containers this was reduced to just over 11 hours.

## Conclusions

The current work presents an update to the optical-propagation plugin for the latest GAMOS release. The features of this Monte Carlo package, namely radiation-induced light signal tracking, have a range of applications from radiation dosimetry to metabolic sensing and therapy, as will be demonstrated in the following chapters. The updates described here are provided in a stand-alone package based on both virtual machine and container architectures to allow for future portability. A method is described for deploying the container-based system using an on-demand cloud infrastructure which can be used to compare models with a range of input parameters, where the resulting time to obtaining can be drastically lower than a sequential model would require. Additionally, this methodology deployment strategy could easily be adapted to fit many other script-base MC packages.

## Acknowledgements

This work was funded by NIH grant R01 EB024498 and R01 EB023909, as well as by a National Science Foundation Graduate Research Fellowship.

## Supplementary Material

1. Additional GAMOS 6.1 optical plugin information and VM files: https://sites.dartmouth.edu/optmed/research-projects/monte-carlo-software/

2. GAMOS 6.1 GamosCore source code including tissue optics plugin: https://github.com/ethanlarochelle/GamosCore/tree/6_1

3. GAMOS 6.1 Optical plugin validation and example scripts: https://github.com/ethanlarochelle/GAMOS_examples

4. GAMOS 6.1 Cloud infrastructure notebooks (Interactive environment available upon request via CodeOcean): https://github.com/ethanlarochelle/gamos-cloud

5. GAMOS 6.1 optical plugin container image: https://hub.docker.com/r/ethanlarochelle/gamos_6_1_tissue_optics